\documentclass[twocolumn]{aastex631}
\usepackage{color} 
\usepackage{tabularx} 
\usepackage{epsfig}
\usepackage{amsmath} 
\usepackage{amssymb} 
\usepackage{bm}
\usepackage{graphicx}
\usepackage{wasysym}
\usepackage{multirow}
\usepackage{threeparttable}
\definecolor{purp}{HTML}{8904B1}
\newcommand{\rsun}{R$_\odot$}
\newcommand{\be}{\begin{equation}}
\newcommand{\ee}{\end{equation}}
\newcommand{\dd}{{\rm d}}

\shorttitle{Solar convection-zone changes}
\shortauthors{S. Basu}

\begin{document}

\title{Evidence of solar-cycle related structural changes in the solar convection zone}

\author[0000-0002-6163-3472]{Sarbani Basu}
\affil{Department of Astronomy, Yale University, New Haven, CT 06520, USA}
\email{sarbani.basu@yale.edu}

\begin{abstract}
While it has been relatively easy to determine solar-cycle related changes in solar dynamics, determining changes in structure
in the deeper layers of the Sun has proved to be difficult. By using helioseismic data obtained over two solar cycles, and sacrificing resolution in favour of lower uncertainties, we show that there are significant changes in the
solar convection zone, and perhaps even below it. Using MDI data, we find a relative squared sound-speed difference of $(2.56\pm 0.71)\times 10^{-5}$ at the convection-zone base between the maximum of solar Cycle~23 and the minimum between Cycles~23 and 24. The squared sound-speed difference for the maximum of Cycle~24 obtained with HMI data is $(1.95\pm 0.69)\times 10^{-5}$. GONG data support these results. We also find that the sound speed in the solar convection zone decreases compared to the sound speed below it as the Sun becomes more active. We find
evidence of changes in the radial derivative of the sound-speed difference between the solar minimum and other epochs at the base of the convection zone implying possible small changes in the position of the convection-zone base, however, the results are too noisy to make any definitive estimates of the change.
\end{abstract}

\section{Introduction}
\label{sec:intro}

Solar frequencies, and frequency splittings, change with solar activity. This has been known for
decades \citep{woodard, pere, yvonne1990,libbrecht1990} and is one of the earliest, and most robust, findings of
helioseismology. The frequency changes were shown to be correlated with the average line-of-sight magnetic field over the solar surface \citep{woodard1991ApJ}; this was confirmed later using much longer data sets \citep[e.g.][]{howe2002}. The correlation between frequency shifts and activity is also seen for other solar activity proxies \citep[etc.]{kiran,howe2018}.

The solar-cycle dependent changes in solar oscillation frequencies have been successfully used to determine activity-related changes in solar dynamics (see review by \citealt{rachelLRSP} for Cycle~23 results, subsequent results can be found in \citealt{antiabasu2013, howeetal2013, rachel2018, sasha2019, basu2019}, etc.) While changes in solar dynamics have been quite easy to detect, detecting changes in solar structure has been very difficult. 

Solar dynamics, and changes thereof,  are determined from frequency splittings, while the structure is determined from the frequencies themselves. Solar-cycle dependent frequency changes are predominantly a function of frequency, with the changes increasing with frequency. It is reminiscent of the ``surface term,'' i.e., changes or differences in frequencies caused by near-surface changes or differences. The apparent lack of a degree-dependence in the frequency shifts {has} led to the view that
solar-cycle related changes in the Sun are confined to a thin layer close
to, or even above, the surface \citep[][etc.]{libbrecht1990,goldreich1991,evans,nishizawa}. Other studies seemed to confirm this \citep[][etc.]{ rachel1999, sbhma2000}. Despite this, there have been attempts to look for changes deeper inside the Sun. 

The solar dynamo is believed to be situated in the tachocline which is more or less coincident with the base of the solar convection zone \citep{kosovichev, basu1997, paulchar, basu2019}. 
There has been an expectation that the large magnetic fields generated there will be visible as changes in solar structure close to the convection-zone base. Consequently, there have been attempts, with various degrees of success {in detecting such changes}
\citep[][etc.]{dziem2000, basu2002, antonio2002,chou2005}. 
\citet{baldner} did a principal component analysis of the helioseismic data available until May 2007 to find a change in sound speed at the base of the convection {zone;} the results at other radii were not statistically significant. None of the other efforts has shown a clear, unambiguous pattern of changes that are correlated with solar activity. Attempts to determine changes in relatively shallow layers were, however, successful. \citet{baldner} showed that there was evidence of changes in the outer 10--15\% of the Sun, but at low latitudes, while \citet{basu2007} found evidence of change even at high latitudes, but their work was only sensitive to the outer 2\% of the Sun. \citet{anna} examining signatures caused by the acoustic glitch in the He{\sc ii} ionization zones detected a solar cycle-dependent {change;} this was subsequently confirmed by \citet{courtney} using solar frequencies for Cycles~23 and 24, but while the result was obtained with data sensitive to changes at all latitudes, the change was localized to a shallow layer. 

\citet{courtney} showed that the use of two solar cycles worth of data helps in demonstrating whether or not there is a solar-cycle dependent change when the changes are small. In this paper, we use helioseismic data for two solar cycles to examine whether or not there are changes in the deeper layers of the Sun. We do a differential analysis, i.e., we analyze
frequency differences between different epochs of the two solar cycles and the solar minimum between Cycles~23 and 24. We perform solar structure inversions, however, unlike usual inversions, we sacrifice radial resolution in order to reduce uncertainties in the results. 

The rest of the paper is organized as follows: We describe the data in Section~\ref{sec:data}, the frequency differences are examined in Section~\ref{sec:freqs}, the inversion method and possible systematic errors are
described in Section~\ref{sec:inv}, we present our results in Section~\ref{sec:res}. We discuss the implications of our results in Section~\ref{sec:conclu}.

\section{Data used}
\label{sec:data}

For this work, we use solar oscillation frequencies obtained by the ground-based  Global
Oscillation Network Group (GONG) \citep{hill1996} and the space-based the Michelson Doppler Imager
(MDI) onboard the Solar and Heliospheric Observatory spacecraft \citep{mdi} and the Helioseismic and Magnetic Imager (HMI) \citep{hmi}  onboard the Solar Dynamics Observatory. 

The GONG data we use cover a period from 1995.05.05 to  2020.05.18. The data
are designated by GONG ``months'', each ``month'' being 36 days long. Solar oscillation
frequencies and splittings of sets starting GONG Month~2 are obtained using 108-day
(i.e., 3 GONG months) time-series. There is an overlap of 72 days between
different data sets, i.e., GONG Month~2 frequencies were obtained from data of
GONG Months~1, 2 and 3, those for GONG~3 from GONG Months~2, 3 and 4, etc. We use data for GONG Months~2 to 254. While the
overlapping sets are useful in looking at time variations, they can bias statistics, consequently, all statistical inferences are drawn from non-overlapping data sets. These data sets are publicly available from GONG archives. 

Data from  MDI cover the period from 1996.05.01 to April 24, 2011.04.24. Solar oscillation
frequencies and splittings for these data are obtained from 72-day time series
and the sets have no overlap in time. HMI started obtaining data on 
2010.04.30 and these are also obtained from 72-day time series. We use all MDI data, and all HMI sets up to set 10000, which has an end date of 2020.07.30. These sets are available to all researchers through the Joint Science Operations Centre of the Solar Dynamic Observatory.

We use the 10.7 cm flux as a measure of solar activity \citep{tapping2, tapping}. These data are also public. \footnote{Data are available from\\
 http://www.spaceweather.gc.ca/solarflux/sx-en.php,\\ and
https://omniweb.gsfc.nasa.gov/form/dx1.html}.

For the purpose of looking at time variations and solar-activity related changes, we use GONG data as one set and combine the
MDI and HMI data to form another set --- if both sets show the same trends, we can be confident that we are not seeing an instrumental effect. Since we are doing a differential study, we choose one reference frequency set each for the ground-based and space-based data. In each case, we choose sets from the minimum between Cycle~23 and Cycle~24.
For the GONG sets, we use the averaged frequencies of GM~133, 136 and 139 as the reference set. These are non-overlapping sets and have an average 10.7 cm flux of 68.1 SFU\footnote{1 solar flux unit or SFU is $10^4$ Jansky $ i.e.,\\ 10^{-22}$ W~m$^{-2}$~Hz$^{-1}$}.
For the space-based set, we use a specially created 720-day data set spanning 2007.07.27 to 2009.07.16 as the reference set. This set has an average 10.7 cm flux of 69.3 SFU

\section{A clue to cycle-dependent structure changes}
\label{sec:freqs}

If the solar-cycle related changes of frequencies were a result only of changes localized to the very shallow, near-surface layers of the Sun, the changes would be a function of frequency alone, at least at low and intermediate degrees; any degree differences can be accounted for by the differences in the inertia of the modes. {As in} the case of all surface terms, it is usual to fit an \textit{ad hoc}, slowly varying function of frequency to model the solar-cycle related change.

However, the form of the surface term need not be completely arbitrary. \citet{libbrecht1990} showed that the shifts that they found could be modelled as the cube of the frequency divided by the mode inertia. This was given some theoretical justification by \citet{goldreich1991} who showed that a photospheric perturbation could produce such a term. \citet{gough1990} proposed an additional term, one that depends on the inverse of the frequency, again scaled by the mode inertia. The cubic term
could arise from the presence of some factor that affects sound speed but not density, such as magnetic
flux tubes. The inverse term would reflect factors such as a change in the pressure scale height near the surface.
\citet{bg14} first used this to describe the frequency dependence of the surface term in low-degree solar and asteroseismic data. 

\begin{figure}
\epsscale{1.00}
\plotone{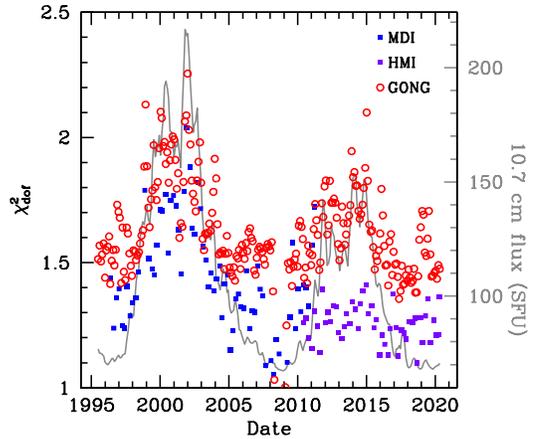}
\caption{ The variation of the reduced $\chi^2$ obtained by fitting Eq.~\ref{eq:bg14} to the frequency differences are plotted as a function of time. The grey curve in the background is the 10.7 cm flux whose values can be read from the axis on the right. 
}
\label{fig:chi2}
\end{figure}

The \citet{bg14} parameterization is given by:
\be
\delta\nu=\frac{1}{E_{nl}}\left[ a_{-1}\left(\frac{\nu}{\nu_{\rm ac}}\right)^{-1} + a_3\left(\frac{\nu}{\nu_{\rm ac}}\right)^{3}\right],
\label{eq:bg14}
\ee
where, $E_{nl}$ is the inertia of a model of degree $l$ and order $n$, and $\nu_{\rm ac}$ is the acoustic cut-off frequency. \citet{rachel} showed that this can be applied to both low-degree and intermediate degree solar modes. What had not been examined directly in that work is how the fits behave as a function of time, i.e., how the reduced $\chi^2$ behaves as a function of time. The reduced $\chi^2$ for the frequency differences between the individual GONG sets and the GONG reference set, as well as the frequency differences of the MDI and HMI sets with respect to the MDI reference set, are shown in Fig.~\ref{fig:chi2}. What is clear immediately is that there is a solar-cycle-dependent change in the reduced $\chi^2$ and that the $\chi^2$ is largest at the solar maxima.

\begin{figure}
\epsscale{1.00}
\plotone{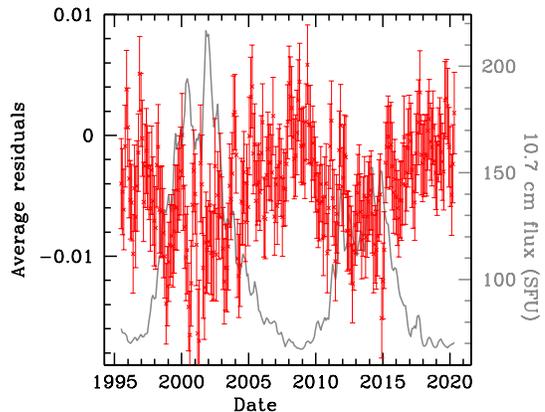}
\caption{ The time-variation of the average residuals for modes with lower turning points between $0.7125\pm 0.0125$\;\rsun\ obtained from the fits to Eq.~\ref{eq:bg14}. Only GONG data are shown for the sake of clarity. The grey curve in the background is the 10.7 cm flux whose values can be read from the axis on the right. 
}
\label{fig:resid}
\end{figure}

The variation of the reduced $\chi^2$ is an indication that there is more to the solar cycle-dependence of the
frequencies differences than just a contribution from the near-surface layers. In Fig.~\ref{fig:resid} we show how  the
average residuals for modes with lower-turning points between $0.7125\pm 0.0125$\rsun vary as a function of time. Since modes that have their lower turning point within any radius range can have different frequencies --- recall that the classical lower turning point of a mode is $r_t=c(r_t)\sqrt{l(l+1)}/\omega$, where $c(r_t$) is the sound speed at $r_t$ and $\omega=2\pi\nu$ --- a frequency-dependent solar-cycle related change alone cannot explain a solar-cycle related change in the frequency of modes with a given lower turning point. Fig.~\ref{fig:resid} points to changes in the structure around $0.7125\pm 0.0125$\rsun, which means we cannot rule out changes at other radii either. We look for changes in the internal sound speed by inverting the frequency difference between the different data sets and the reference sets.

\section{Inversions: Methods and systematic errors}
\label{sec:inv}

Inversions for solar structure start with the linearized form of the oscillation equations obtained using the variation principle:
\be
\frac{\delta\nu_i}{\nu_i}=\int K^i_{c^2,\rho}\frac{\delta c^2}{c^2} \dd r+\int K^i_{\rho, c^2}\frac{\delta\rho}{\rho} \dd r+ \frac{F(\nu_i}{E_i},
\label{eq:inv}
\ee
where $\delta\nu_i/\nu_i$ is the relative frequency difference of the $i$th mode, $\rho$ is the density, $K^i_{c^2,\rho}$ and $K^i_{\rho, c^2}$ are known functions of a reference solar model, $F_(\nu_i)$ is a slowly varying function function of frequency to account for the surface term, and $E_i$ is the inertia of the $i$th mode. The aim of inversions is to determine $\delta c^2/c^2$ or $\delta\rho/\rho$ given a set of
$\delta\nu/\nu$.

There are two primary ways of inverting Eq.~\ref{eq:inv} that have different aims, these are
The Regularised Least Squares (RLS) method, and the method of Optimally Localised Averages (OLA). 
RLS aims to find the $\delta c^2/c^2$ and $\delta\rho/\rho$ profiles that give the best fit to the data (i.e., give the smallest residuals) while keeping uncertainties small; the aim of OLA is not to fit the data at all, but to find linear combinations of the frequency differences in such a way that the corresponding combination of kernels, the \textit{averaging} or \textit{resolution} kernel, provides a localised average of the unknown function, again while keeping uncertainties small.  Details of how these techniques are implemented can be found in \citet{basu2016}. OLA and RLS inversions are complementary in nature (Sekii 1997) and inversions can be trusted if both inversion techniques return the same results.

There are two variants of OLA, the original version, which in helioseismic literature is referred to as multiplicative OLA (MOLA) because a weighting function used in the construction of the averaging kernels are multiplied with the mode kernels, and subtractive OLA \citep[SOLA;][]{sola}, where the difference between the averaging kernels and a target kernel. SOLA has the advantage that the target kernels can be selected for specific inversions, such as the derivatives of the function rather than the function itself \citep{derivative}. Details of how SOLA and MOLA implementations differ can be found in \citet{basu2016}.

\begin{figure}
\epsscale{1.0}
\plotone{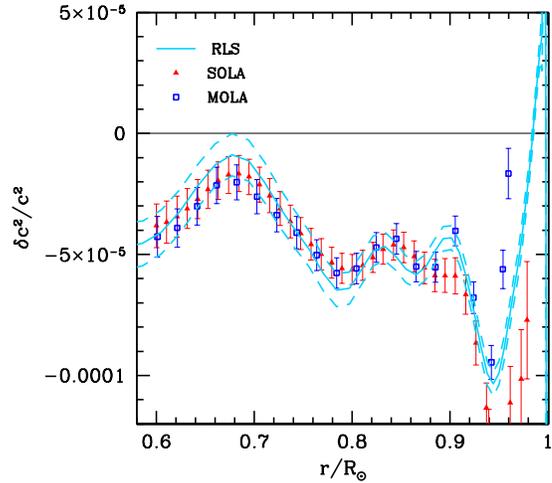}
\caption{ Inversion results for the difference in frequencies between MDI set 3232 and the 2year MDI set obtained using three different inversion methods. Note that the results for the different methods agree at radii below about {0.88}  \rsun. Although not shown in the figure, the match between the results deteriorates below about 0.45\rsun.
}
\label{fig:soln}
\end{figure}

We use SOLA of the results obtained in this paper since it is faster and allows us to directly invert for
derivatives. We use mode kernels from Model~S \citep{modelS}; our tests have shown that the kernels
from other standard solar models give identical results. In Fig.~\ref{fig:soln} we show the inversion results for MDI set 3232 with respect to the 2-yr MDI set obtained with all three methods, 
MOLA, SOLA and RLS, to determine the region where we can trust the inversion results. As can be seen, all three methods agree for radii below about $0.88$\rsun. We have not shown the results deeper than 0.6\rsun\ because the RLS results there are linear, implying that they are dominated by the regularization term rather than the data, and hence not to be trusted; they also deviate from both SOLA and MOLA results.

\begin{figure}
\epsscale{1.00}
\plotone{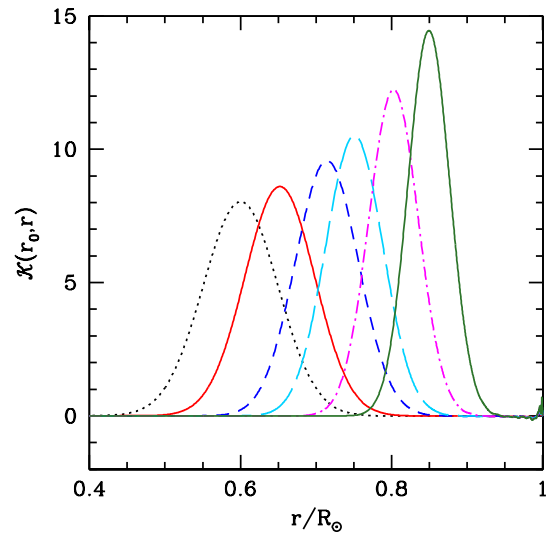}
\caption{ A few averaging kernels (i.e., resolution kernels) for the SOLA inversions shown in Fig.~\ref{fig:soln}. Note that they have very little structure away from the peak.
}
\label{fig:avker}
\end{figure}

Since there is a trade-off between the resolution of the inversions and the uncertainty in the results, we sacrificed resolution in order to increase the precision of the results. The averaging kernels at a few different positions obtained for the SOLA inversions are shown in Fig.~\ref{fig:avker}. As can be seen, the results at most radii will be correlated.

A possible source of systematic error is the structure of the averaging kernel far from the target radius. This is caused mainly because of the lack of high degree modes in the data set. In particular, the worry is that since the averaging kernels are not strictly zero close to the surface, if there is a large solar-cycle related change in the surface, this would leak into the results at other radii giving a false impression about the deeper layers.  Recall that the SOLA inversion results are in effect \citep{basu2016}
\be
\left(\frac{\delta c^2}{c^2}\right)_{\rm inv}=\sum_i c_i \frac{\delta\nu_i}{\nu_i} = \sum_i c_i \int K^i_{c^2,\rho}\frac{\delta c^2}{c^2}\dd r,
\label{eq:result}
\ee
assuming that $\sum_i c_i \int K^i_{\rho,c^2}{\delta \rho/\rho}\dd r$ and $\sum_i c_i F(\nu_i/E_i$,  are small; $c_i$ are the inversion coefficients. The term $\sum_ic_i K^i_{c^2,\rho}$ is nothing but $\mathcal{K}$, the averaging kernel, and $\sum_ic_i K^i_{\rho,c^2}$ is usually called the `cross term' kernel. Since the inverted value of the
sound speed at a radius $r_0$ is in effect $\int \mathcal{K}(r_0;r)\delta c^2/c^2 \dd r$, any structure in the averaging kernel away from $r_0$ will contribute to the solution. As can be seen from Fig.~\ref{fig:avker}, the averaging kernels are not zero close to the surface.
\begin{figure}
\epsscale{1.0}
\plotone{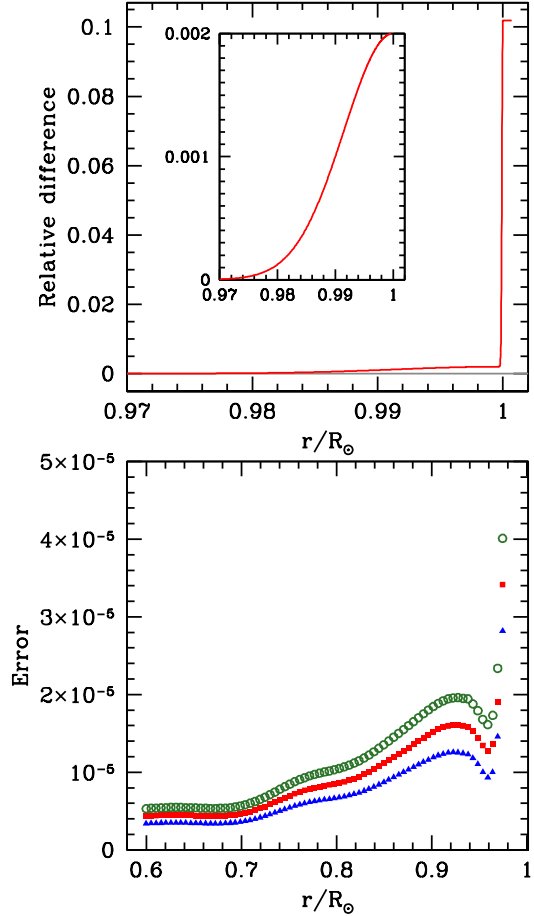}
\caption{ Possible systematic error in solution caused by near-surface structure in the averaging kernels. Top: the assumed difference in $\delta c^2/c^2$, about 10\% at the surface, with the inset showing the difference near the helium ionization zone. Bottom: The systematic error at different radii. The red squares assume that only $\delta c^2/c^2$ at the surface matters; the blue triangles assume that there is also a relative difference in density that is identical both in magnitude and sign to the assumed $\delta c^2/c^2$, while the green circles assume that the density differences have identical magnitude, but are opposite in sign to that of the assumed $\delta c^2/c^2$. 
}
\label{fig:error}
\end{figure}

To try and quantify the effect of the surface structure of the averaging and cross-term kernels on the results, we assume a large change in structure (10\%) at the surface (see the top panel of Fig.~\ref{fig:error}, and we also assume that there are changes as deep as the second helium ionization zone since there are indirect detections of change in structure in those layers \citep{courtney}. We then determine the effect of this change on the solutions at different radii. We show the results in the lower panel of Fig.~\ref{fig:error}. We show the results assuming that there is a difference only in sound speed, a change in both sound speed and density with both changing the same way, and in one case where the changes in sound speed and density have the same magnitude but opposite sign. As can be seen in the figure, the systematic error because of unresolved surface changes of 10\% is less than a part in $10^5$ for radii below 0.8\rsun\ and rises thereafter. This will need to be kept in mind while interpreting the results {at different radii. Clearly, for us to claim that we see changes in  $c^2$,  they need to be greater than 1 part in $10^5$.}

\section{Results}
\label{sec:res}

\subsection{Changes in sound speed}
\label{subsec:sound}

\begin{figure*}
\epsscale{0.70}
\plotone{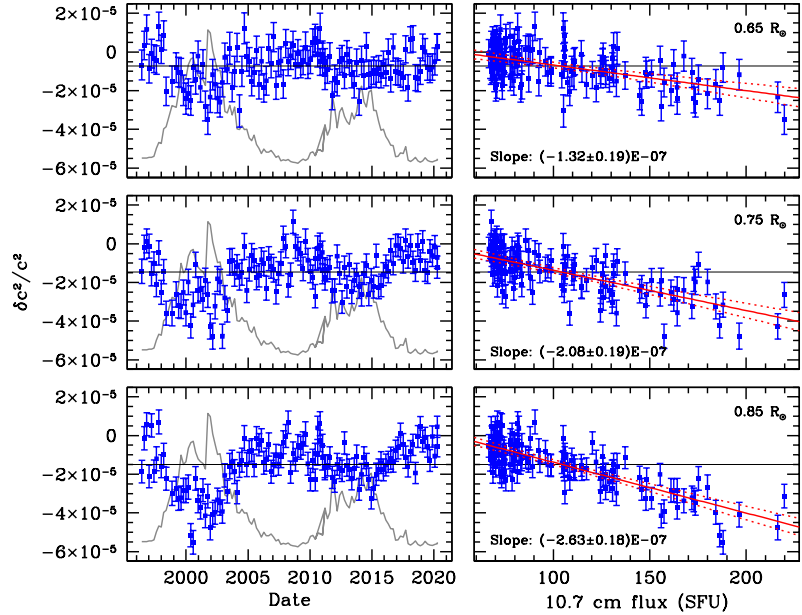}
\caption{ Inversion results for MDI and HMI data sets at three radii. \textbf{Left}: Results plotted as a function of time, with the curve in the background showing the 10.7 cm flux plotted on a scale from 0 to 250 (not shown). The horizontal line in each panel is the weighted average of the results. Note that while the results at 0.65\rsun\ and 0.85\rsun\ are independent (see Fig.~\ref{fig:avker}), the result at 0.75\rsun\  has an overlap with both 0.65 and 0.85\rsun. \textbf{Right:} The same results plotted as a function of the 10.7~cm flux. The horizontal line is again the weighted average. The red line shows a weighted linear least-squares-fit to the points with the 95\% confidence level shown as the dotted lines.
}
\label{fig:space}
\end{figure*}

\begin{figure*}
\epsscale{0.70}
\plotone{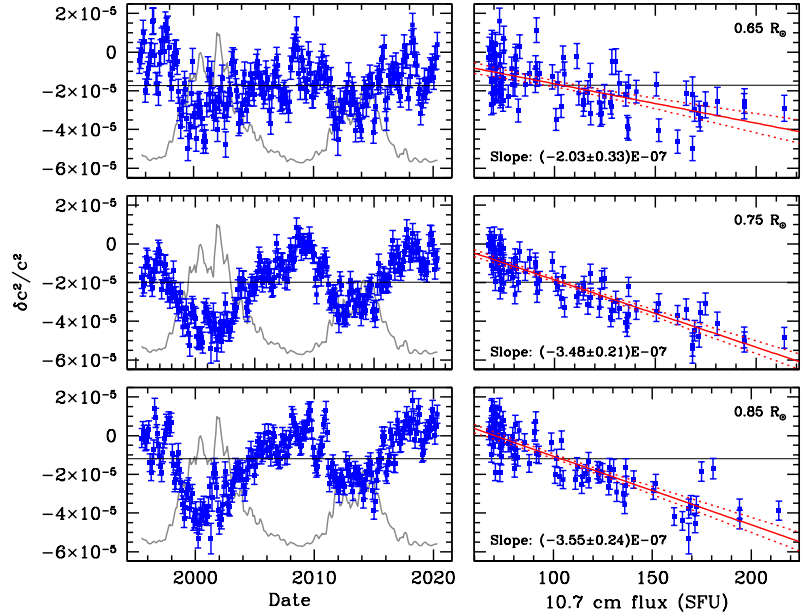}
\caption{ The same as Fig.~\ref{fig:space}, but for GONG data. Note that while we plot all sets against time, we only plot non-overlapping sets against the radio flux to avoid exaggerating the correlation.
}
\label{fig:ground}
\end{figure*}

In Figs.~\ref{fig:space} and \ref{fig:ground} we show the inversion results obtained with {space based} data (MDI+HMI) and GONG respectively at three different radii. The results are plotted both as a function of time and of the 10.7 cm flux. In the case of GONG data, the results of all sets are plotted against time, but only non-overlapping sets are plotted against the radio flux.
It can be seen very clearly that there is a time variation in the result. Note that the results are with respect to the minimum between Cycle~23 and 24, and the results show that the sound speed is lower at high activity than at low activity, {particularly} in the convection zone. The changes are large enough that they cannot be explained as systematic error due to changes in the
near-surface layers. Visually, the GONG results look more significant, but that is probably {because} the sets overlap in time. The figures appear to show that there is a variation in sound speed even below the base of the convection zone, but as can be seen from Fig.~\ref{fig:avker}, the averaging kernel for 0.65\rsun\ samples the convection zone too.

As expected from the time variation,  the changes are well correlated with the 10.7 cm flux at all radii. The correlation coefficients and their statistical significance are listed in Table~\ref{tab:corr}. Note that the lower the $p$ value, the more the significance of the slope; a  $p$-value of 0.05 corresponds to a 95\% confidence level. The slopes of the relation between the changes and the 10.7~cm flux obtained for the space-based and ground-based data do not completely agree --- GONG results are steeper. This is most likely due to the fact that the modes sets used are different for GONG and MDI+HMI, and also that for the GONG sets we have used a reference set of data that spans slightly less than a year, while for MDI+HMI we subtract out a 2-year set.  What is important though, is that both sets show that there is an activity-related change in structure. We also note that the change seems to be larger within the convection zone than below it, but as mentioned earlier the results below the convection zone have a substantial contribution from those in the convection zone.

\begin{table}
  \centering
  \label{t:spec}
  \tabletypesize{\footnotesize}
  \caption{Correlations and statistical significance}
   \scriptsize
  \begin{tabular}{lccccc}
    \toprule
    Quantity & Radius & Pearson & Spearman & $t$ value & $p$ value \\
             & (\rsun) & coeff. & coeff. &             &  \\
    \hline
     & \multicolumn{5}{c}{MDI+HMI}\\
     \hline
     $\delta c^2/c^2$ & 0.650 & $-0.54$  & $-0.46$ & $-5.83$ & $< 0.00001 $\\
     $\delta c^2/c^2$ & 0.750 & $-0.71$  & $-0.69$ & $-10.63$ & $< 0.00001 $\\
     $\delta c^2/c^2$ & 0.850 & $-0.80$  & $-0.74$ & $-12.14$ & $< 0.00001 $\\
     $\Delta(\delta c^2/c^2)$ & 0.713 & $0.37$  & $0.33$ & $3.91$ & $ 0.00015$\\
     $\frac{\rm d}{{\rm d} r}(\delta c^2/c^2)$ & 0.613 & $-0.02$  & $-0.04$ & $-0.48$ & $ 0.64 $\\
     $\frac{\rm d}{{\rm d} r}(\delta c^2/c^2)$ & 0.713 & $-0.37$  & $-0.35$ & $-4.21$ & $ 0.00006 $\\
     $\frac{\rm d}{{\rm d} r}(\delta c^2/c^2)$ & 0.813 & $-0.28$  & $-0.18$ & $-2.06$ & $ 0.04 $\\
     $(\delta c^2/c^2)_{\rm alt}$ & 0.750 & $-0.61$  & $-0.60$ & $-8.17$ & $< 0.00001 $\\
    \hline
    & \multicolumn{5}{c}{GONG}\\
    \hline
    $\delta c^2/c^2$ & 0.650 & $-0.58$  & $-0.72$ & $-6.43$ & $< 0.00001 $\\
     $\delta c^2/c^2$ & 0.750 & $-0.88$  & $-0.88$ & $-17.29$ & $< 0.00001 $\\
     $\delta c^2/c^2$ & 0.850 & $-0.85$  & $-0.83$ & $-13.74$ & $< 0.00001 $\\
     $\Delta(\delta c^2/c^2)$ & 0.713 & $0.40$  & $0.37$ & $3.60$ & $ 0.00027$\\
     $\frac{\rm d}{{\rm d} r}(\delta c^2/c^2)$ & 0.613 & $-0.24$  & $-0.23$ & $-2.81$ & $ 0.33 $\\
     $\frac{\rm d}{{\rm d} r}(\delta c^2/c^2)$ & 0.713 & $-0.45$  & $-0.50$ & $-4.71$ & $ <0.00001 $\\
     $\frac{\rm d}{{\rm d} r}(\delta c^2/c^2)$ & 0.813 & $-0.14$  & $0.05$ & $-0.45$ & $ 0.016 $\\
     $(\delta c^2/c^2)_{\rm alt}$ & 0.750 & $-0.88$  & $-0.89$ & $-18.01$ & $< 0.00001 $\\
    \toprule
 \end{tabular}
  \label{tab:corr}
\end{table}

\subsection{Differences below and above the convection zone}
\label{subsec:below}

\begin{figure}
\epsscale{1.00}
\plotone{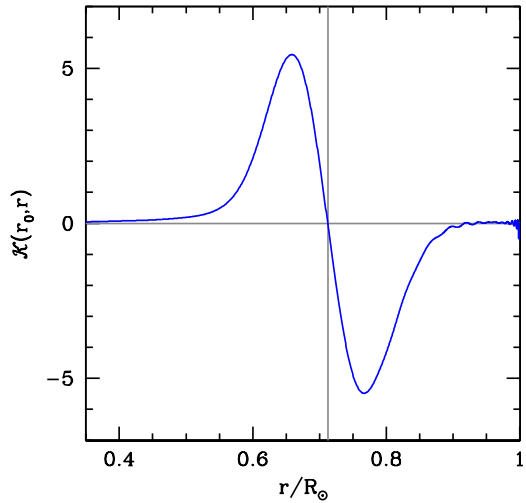}
\caption{ The averaging kernels for inversions to determine the difference above and below the convection-zone base. The vertical line marks the position of the convection-zone base.
}
\label{fig:cz}
\end{figure}

\begin{figure}
\epsscale{1.20}
\plotone{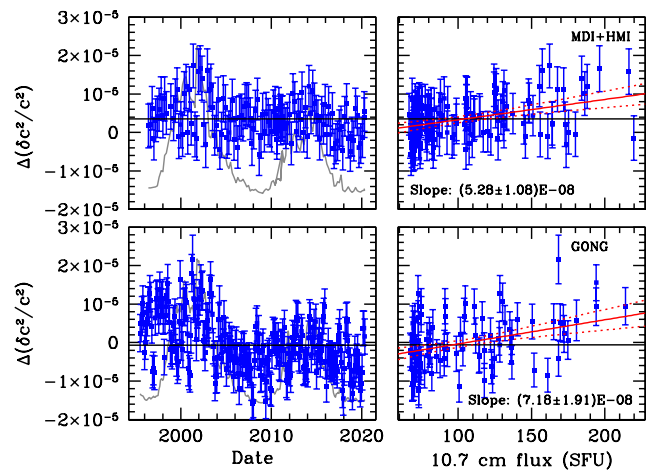}
\caption{ Inversion results for the difference in sound speed below and above the convection-zone base. As with previous figures, although all GONG results are plotted as a function of time, only results from non-overlapping data sets are plotted again the 10.7 cm flux.
}
\label{fig:tach}
\end{figure}

The results in Figs.~\ref{fig:space} and \ref{fig:ground}point to differences below and above the convection zone, however, as mentioned earlier, the difference below the convection zone are affected by differences above, and hence merely subtracting the results will not give us the true difference. We make use of the fact that for SOLA inversions we can define a suitable target averaging kernel that will give us the quantity we seek, in this case, the difference in the results at two locations. The averaging kernel obtained for directly inverting for the difference in sound speed below and above the base of the convection zone is shown in Fig.~\ref{fig:cz}. Note that we get a clean averaging kernel with no overlap below and above the base of the convection zone. 
The results of the inversion are shown in Fig.~\ref{fig:tach}. There is indeed a small, time-varying difference between the layer below and above the convection zone, with the sound speed in the convection zone being lower than that below it when the Sun is active. The differences are {small;} however, the systematic error in the results for an artificial profile like that in Fig.~\ref{fig:error} is at most 3 parts in $10^6$, i.e., smaller than the changes seen. 

\subsection{Is there a change in the position of the convection-zone base?}
\label{subsec:czbase}

\begin{figure}
\epsscale{1.00}
\plotone{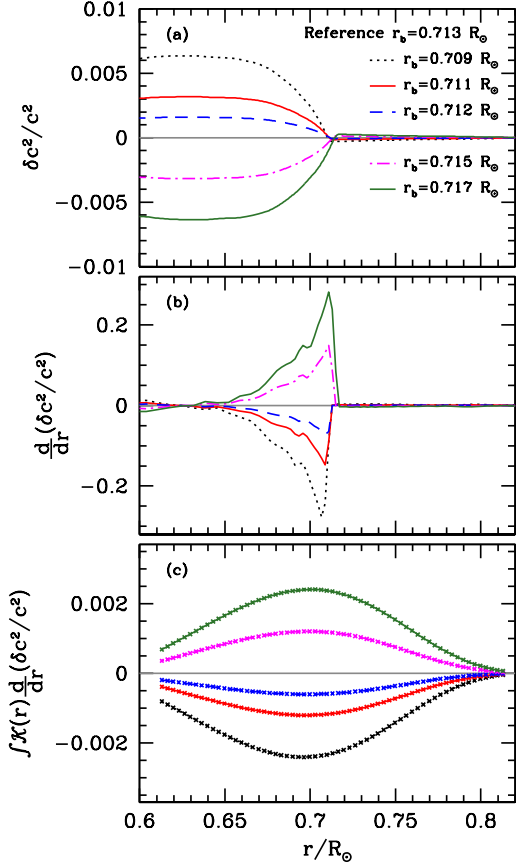}
\caption{ (a) The relative squared sound-speed difference between models with different positions of the base of the convection zone and a model with the convection-zone base at 0.713\rsun. (b) The radial gradient of the $\delta c^2/c^2$ shown in panel (a). (c) Results that would be obtained if we were to invert the frequency differences between the models. These results are obtained by multiplying the averaging kernel for each of the points with the profiles in panel (b). Note that the value is highest at the base of the convection zone and quickly tapers out on each side.
}
\label{fig:interp}
\end{figure}
\begin{figure}
\epsscale{0.90}
\plotone{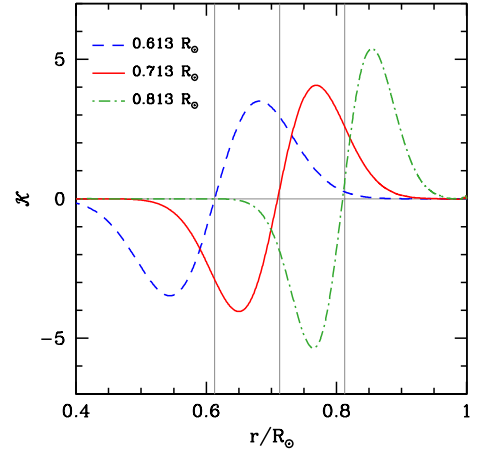}
\caption{ The averaging kernels for inversions to determine the derivative of $\delta c^2/c^2$ at 0.613, 0.713 and 0.813\rsun.
}
\label{fig:avker_deriv}
\end{figure}

The base of the convection zone is marked by a change in the temperature gradient, consequently. the sound-speed difference for two models with different $r_b$, the position of the convection-zone base shows a sharp rise and a steep gradient near $r_b$, as is shown in Fig.~\ref{fig:interp}.

We use the SOLA technique to determine the derivative of $\delta c^2/c^2$  {directly;} these are
shown for three different radii in Fig.~\ref{fig:avker_deriv}. Since the averaging kernels are quite wide, we cannot expect to see the steep gradients that are seen in Fig.~\ref{fig:interp}(b), instead, the derived results will be smaller and smoother, as shown in Fig.~\ref{fig:interp}(c). However, Fig.~\ref{fig:interp}(c) shows that even though the inversions will smooth out the underlying gradient, the obtained difference will be largest near the position of the convection-zone base.

\begin{figure}
\epsscale{1.20}
\plotone{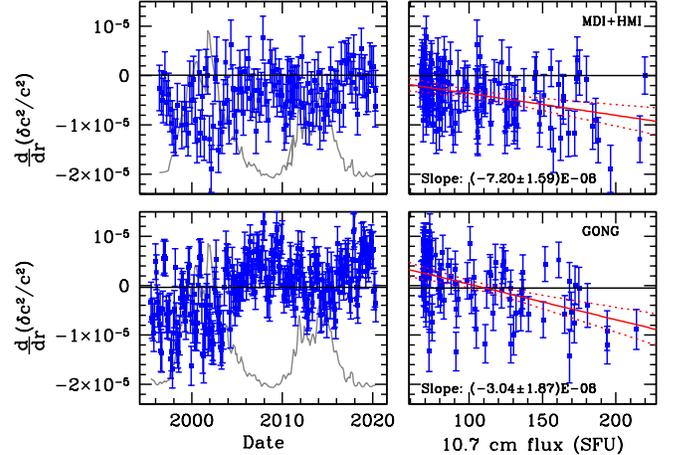}
\caption{ The radial derivative of $\delta c^2/c^2$ at 0.713\rsun\ plotted as a function of time and the 10.7 cm flux.
}
\label{fig:deriv}
\end{figure}

The inversion results at 0.713\rsun (the base of the solar convection zone as per
\citealt{jcd1991}, \citealt{basu1998}) can be seen in Fig.~\ref{fig:deriv}. The results appear to show a variation with time and activity, however, the results are noisy enough that to actually determine the amount of change in the position of the convection-zone base is extremely difficult. We do not show the results at other radii since they are not very statistically significant. Additionally, as seen from Fig.~\ref{fig:interp}, if there is a
change in the position of the convection-zone base, it will be the largest in the region around it.
If we try to interpret Fig.~\ref{fig:deriv}(c) at face value with the aid of Fig.~\ref{fig:interp}, it would appear that the convection zone becomes deeper at activity maximum.

\subsection{Is the surface term fooling us?}

\begin{figure*}
\epsscale{0.70}
\plotone{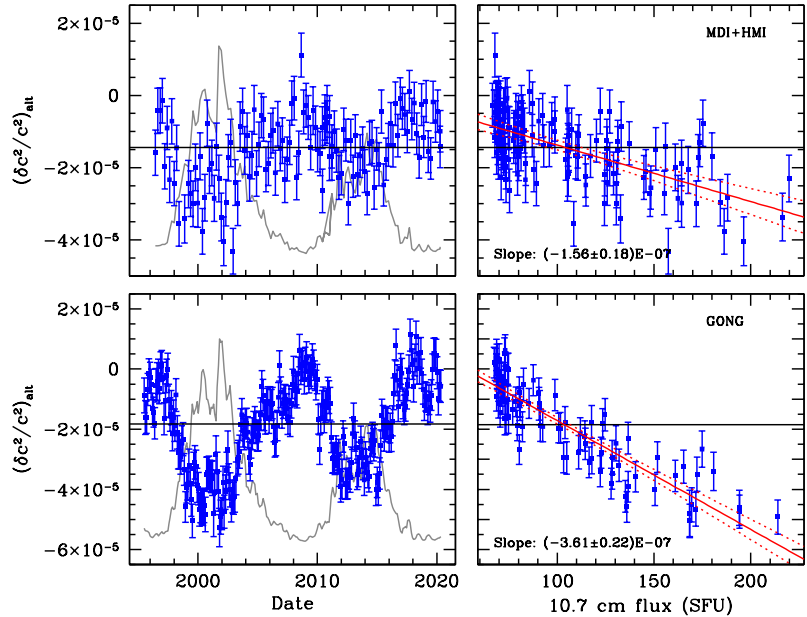}
\caption{ Inversion results at 0.750\rsun\ obtained using an alternative surface term.
The slopes are almost identical to those obtained with the BG14 surface term (Figs.~\ref{fig:space} and \ref{fig:ground}).
}
\label{fig:legpol}
\end{figure*}

Unlike earlier attempts in determining solar cycle-related structural changes, we have used a very specific form of the surface term, that given in Eq.~\ref{eq:bg14}. Thus one could wonder whether the results are simply an artefact of using that form of the surface term, and perhaps that form does not really account for all near-surface effects. To examine whether this is indeed the case, we repeated the inversions with a conventional surface term. The results at
0.75\rsun\ are shown in Fig.~\ref{fig:legpol}, and as can be seen, we see the same type of time variation and the slope of the change against the 10.7 cm radio flux is almost identical. Thus the form of the surface term is not causing the time-variation that we find.

\section{Discussion}
\label{sec:conclu}

We have used two solar cycle's worth of helioseismic data to show that there is indeed
a solar-cycle related change in structure, particularly in the convection zone. \citet{courtney} had to use indirect means, we observe the changes more directly by using conventional inversions, however, using current mode sets, we can only probe changes between about 0.6 and 0.88 \rsun. Although we started the study using the \citet{bg14} formulation of the surface term, we find that the time dependence remains when we use the convection forms of the surface term. The changes, although small, are statistically significant.

\citet{baldner} had found a change $\delta c^2/c^2$ of $(7.23\pm 2.08) \times 10^{-5}$ near the base of the
convection zone between the solar minimum and maximum. The
changes they find are localized at $0.712^{+0.0097}_{-0.0029}$ \rsun.  In
contrast we get a change of $(2.56\pm 0.71)\times 10^{-5}$ using MDI data and $(3.99\pm 0.65)\times 10^{-5}$ using GONG data. The discrepancy in the results is most likely due to differences in the radial resolution of the two sets of inversions; in this work, we sacrificed resolution to lower the uncertainties in the solution our uncertainties are smaller nearly a factor of 3 compared with that of \citet{baldner}. Unlike the highly oscillatory nature of the \citet{baldner} solution,  our differences are smooth at all radii (see Fig.~\ref{fig:soln}) which again points to the effects of the radial resolution of the inversions.

The changes in solar dynamics between Cycles~23 and 24 were quite different \citep{rachel, basu2019}. In particular, changes in the tachocline did not show the same activity-related change in the two cycles --- the change in the rotation rate across the tachocline was different for the same value of the 10.7 cm flux. Given the uncertainty in our results, we are not able to examine cycle-to-cycle changes between the response of internal sound speed to solar activity. A principal component analysis of the two cycles,  done separately, might be able to detect cycle-to-cycle changes since there is evidence that the frequency response to solar activity was different in the two cycles \citep{broomhall2015}.

Our results show that as solar activity increases, the sound speed in the region above the base of the convection zone, i.e., the tachocline region, decreases compared with that below the convection-zone base. This result is consistent with the assumption that the sound-speed changes are a result of magnetic fields. 

By directly inverting for the gradient of the sound-speed differences, we have shown that there is a change
in the gradient with the solar cycle. The change is most significant at the base of the convection zone. Interpreted
most simply, this implies a change in the position of the convection-zone base. However, with currently available
frequencies, finding the change at any given epoch is difficult. There had been earlier attempts to determine if the position of the convection-zone base showed any change without much success \citep{basu1999soho}, however, those  used a very limited data set (for instance, \citealt{basu1999soho} only had access to data from 1995 -- 1998), and did not use a differential method.  

The results are intriguing enough to examine the results further. Given that the current data sets are not ideal, the {analyses} need to be repeated with frequencies obtained from time series that are longer; while longer time series will smooth out some of the solar-cycle related changes, overlapping sets will still reveal the changes. Two-year data sets at the solar maxima and two years at the
minima and a few two-year sets in between will help.

\begin{acknowledgments}
This work utilizes data from the National Solar Observatory Integrated Synoptic Program, which is operated by the Association of Universities for Research in Astronomy, under a cooperative agreement with the National Science Foundation and with additional financial support from the National Oceanic and Atmospheric Administration, the National Aeronautics and Space Administration, and the United States Air Force. The GONG network of instruments is hosted by the Big Bear Solar Observatory, High Altitude Observatory, Learmonth Solar Observatory, Udaipur Solar Observatory, Instituto de Astrofísica de Canarias, and Cerro Tololo Interamerican Observatory. This work also used data from the Michelson Doppler Imager on board SOHO. SOHO is a project of international cooperation between ESA and NASA. We also use data from the Helioseismic and Magnetic Imager on board the Solar Dynamics Observatory and acknowledge the HMI consortium for their support. Particular than are due to
 Timothy Larson and Charles Baldner for analyzing the 2-year MDI data set. We would also like to thank the referee for pointing out issues with the text.
\end{acknowledgments}

\facility{GONG, MDI, HMI, Penticton Radio Observatory} 


\begin{thebibliography}{}
\expandafter\ifx\csname natexlab\endcsname\relax\def\natexlab#1{#1}\fi
\providecommand{\url}[1]{\href{#1}{#1}}
\providecommand{\dodoi}[1]{doi:~\href{http://doi.org/#1}{\nolinkurl{#1}}}
\providecommand{\doeprint}[1]{\href{http://ascl.net/#1}{\nolinkurl{http://ascl.net/#1}}}
\providecommand{\doarXiv}[1]{\href{https://arxiv.org/abs/#1}{\nolinkurl{https://arxiv.org/abs/#1}}}

\bibitem[{{Antia} \& {Basu}(2013)}]{antiabasu2013}
{Antia}, H.~M., \& {Basu}, S. 2013, in Journal of Physics Conference Series,
  Vol. 440, Journal of Physics Conference Series, 012018

\bibitem[{{Baldner} \& {Basu}(2008)}]{baldner}
{Baldner}, C.~S., \& {Basu}, S. 2008, \apj, 686, 1349, \dodoi{10.1086/591514}

\bibitem[{{Ball} \& {Gizon}(2014)}]{bg14}
{Ball}, W.~H., \& {Gizon}, L. 2014, \aap, 568, A123,
  \dodoi{10.1051/0004-6361/201424325}

\bibitem[{{Basu}(1997)}]{basu1997}
{Basu}, S. 1997, \mnras, 288, 572

\bibitem[{{Basu}(1998)}]{basu1998}
---. 1998, \mnras, 298, 719, \dodoi{10.1046/j.1365-8711.1998.01690.x}

\bibitem[{{Basu}(2002)}]{basu2002}
{Basu}, S. 2002, in ESA Special Publication, Vol. 508, From Solar Min to Max:
  Half a Solar Cycle with SOHO, ed. A.~{Wilson}, 7--14.
\newblock \doarXiv{astro-ph/0205049}

\bibitem[{{Basu}(2016)}]{basu2016}
---. 2016, Living Reviews in Solar Physics, 13, 2,
  \dodoi{10.1007/s41116-016-0003-4}

\bibitem[{{Basu} \& {Antia}(1999)}]{basu1999soho}
{Basu}, S., \& {Antia}, H.~M. 1999, in SOHO-9 Workshop on Helioseismic
  Diagnostics of Solar Convection and Activity, Vol.~9, 38

\bibitem[{Basu \& Antia(2000)}]{sbhma2000}
Basu, S., \& Antia, H.~M. 2000, Solar Phys., 192, 449

\bibitem[{{Basu} \& {Antia}(2019)}]{basu2019}
{Basu}, S., \& {Antia}, H.~M. 2019, \apj, 883, 93,
  \dodoi{10.3847/1538-4357/ab3b57}

\bibitem[{{Basu} {et~al.}(2007){Basu}, {Antia}, \& {Bogart}}]{basu2007}
{Basu}, S., {Antia}, H.~M., \& {Bogart}, R.~S. 2007, \apj, 654, 1146,
  \dodoi{10.1086/509251}

\bibitem[{{Basu} \& {Mandel}(2004)}]{anna}
{Basu}, S., \& {Mandel}, A. 2004, \apjl, 617, L155, \dodoi{10.1086/427435}

\bibitem[{{Broomhall} \& {Nakariakov}(2015)}]{broomhall2015}
{Broomhall}, A.~M., \& {Nakariakov}, V.~M. 2015, \solphys, 290, 3095,
  \dodoi{10.1007/s11207-015-0728-6}

\bibitem[{{Charbonneau} {et~al.}(1999){Charbonneau}, {Christensen-Dalsgaard},
  {Henning}, {Larsen}, {Schou}, {Thompson}, \& {Tomczyk}}]{paulchar}
{Charbonneau}, P., {Christensen-Dalsgaard}, J., {Henning}, R., {et~al.} 1999,
  \apj, 527, 445

\bibitem[{{Chou} \& {Serebryanskiy}(2005)}]{chou2005}
{Chou}, D.-Y., \& {Serebryanskiy}, A. 2005, \apj, 624, 420,
  \dodoi{10.1086/428925}

\bibitem[{{Christensen-Dalsgaard} {et~al.}(1991){Christensen-Dalsgaard},
  {Gough}, \& {Thompson}}]{jcd1991}
{Christensen-Dalsgaard}, J., {Gough}, D.~O., \& {Thompson}, M.~J. 1991, \apj,
  378, 413, \dodoi{10.1086/170441}

\bibitem[{{Christensen-Dalsgaard} {et~al.}(1996){Christensen-Dalsgaard},
  {Dappen}, {Ajukov}, {Anderson}, {Antia}, {Basu}, {Baturin}, {Berthomieu},
  {Chaboyer}, {Chitre}, {Cox}, {Demarque}, {Donatowicz}, {Dziembowski},
  {Gabriel}, {Gough}, {Guenther}, {Guzik}, {Harvey}, {Hill}, {Houdek},
  {Iglesias}, {Kosovichev}, {Leibacher}, {Morel}, {Proffitt}, {Provost},
  {Reiter}, {Rhodes}, {Rogers}, {Roxburgh}, {Thompson}, \& {Ulrich}}]{modelS}
{Christensen-Dalsgaard}, J., {Dappen}, W., {Ajukov}, S.~V., {et~al.} 1996,
  Science, 272, 1286, \dodoi{10.1126/science.272.5266.1286}

\bibitem[{{Dziembowski} {et~al.}(2000){Dziembowski}, {Goode}, {Kosovichev}, \&
  {Schou}}]{dziem2000}
{Dziembowski}, W.~A., {Goode}, P.~R., {Kosovichev}, A.~G., \& {Schou}, J. 2000,
  \apj, 537, 1026, \dodoi{10.1086/309056}

\bibitem[{{Eff-Darwich} {et~al.}(2002){Eff-Darwich}, {Korzennik},
  {Jim{\'e}nez-Reyes}, \& {P{\'e}rez Hern{\'a}ndez}}]{antonio2002}
{Eff-Darwich}, A., {Korzennik}, S.~G., {Jim{\'e}nez-Reyes}, S.~J., \&
  {P{\'e}rez Hern{\'a}ndez}, F. 2002, \apj, 580, 574, \dodoi{10.1086/343063}

\bibitem[{{Elsworth} {et~al.}(1990){Elsworth}, {Howe}, {Isaak}, {McLeod}, \&
  {New}}]{yvonne1990}
{Elsworth}, Y., {Howe}, R., {Isaak}, G.~R., {McLeod}, C.~P., \& {New}, R. 1990,
  \nat, 345, 322

\bibitem[{{Evans} \& {Roberts}(1992)}]{evans}
{Evans}, D.~J., \& {Roberts}, B. 1992, \nat, 355, 230, \dodoi{10.1038/355230a0}

\bibitem[{{Goldreich} {et~al.}(1991){Goldreich}, {Murray}, {Willette}, \&
  {Kumar}}]{goldreich1991}
{Goldreich}, P., {Murray}, N., {Willette}, G., \& {Kumar}, P. 1991, \apj, 370,
  752, \dodoi{10.1086/169858}

\bibitem[{{Gough}(1990)}]{gough1990}
{Gough}, D.~O. 1990, {Comments on Helioseismic Inference}, ed. Y.~{Osaki} \&
  H.~{Shibahashi}, Vol. 367, 283, \dodoi{10.1007/3-540-53091-6}

\bibitem[{{Hill} {et~al.}(1996){Hill}, {Stark}, {Stebbins}, {Anderson},
  {Antia}, {Brown}, {Duvall}, {Haber}, {Harvey}, {Hathaway}, {Howe}, {Hubbard},
  {Jones}, {Kennedy}, {Korzennik}, {Kosovichev}, {Leibacher}, {Libbrecht},
  {Pintar}, {Rhodes}, {Schou}, {Thompson}, {Tomczyk}, {Toner}, {Toussaint}, \&
  {Williams}}]{hill1996}
{Hill}, F., {Stark}, P.~B., {Stebbins}, R.~T., {et~al.} 1996, Science, 272,
  1292

\bibitem[{{Howe}(2009)}]{rachelLRSP}
{Howe}, R. 2009, Living Reviews in Solar Physics, 6, 1,
  \dodoi{10.12942/lrsp-2009-1}

\bibitem[{{Howe} {et~al.}(2017){Howe}, {Basu}, {Davies}, {Ball}, {Chaplin},
  {Elsworth}, \& {Komm}}]{rachel}
{Howe}, R., {Basu}, S., {Davies}, G.~R., {et~al.} 2017, \mnras, 464, 4777,
  \dodoi{10.1093/mnras/stw2668}

\bibitem[{{Howe} {et~al.}(2018{\natexlab{a}}){Howe}, {Chaplin}, {Davies},
  {Elsworth}, {Basu}, \& {Broomhall}}]{howe2018}
{Howe}, R., {Chaplin}, W.~J., {Davies}, G.~R., {et~al.} 2018{\natexlab{a}},
  \mnras, 480, L79, \dodoi{10.1093/mnrasl/sly124}

\bibitem[{{Howe} {et~al.}(2013){Howe}, {Christensen-Dalsgaard}, {Hill}, {Komm},
  {Larson}, {Schou}, \& {Thompson}}]{howeetal2013}
{Howe}, R., {Christensen-Dalsgaard}, J., {Hill}, F., {et~al.} 2013, in
  Astronomical Society of the Pacific Conference Series, Vol. 478, Fifty Years
  of Seismology of the Sun and Stars, ed. K.~{Jain}, S.~C. {Tripathy},
  F.~{Hill}, J.~W. {Leibacher}, \& A.~A. {Pevtsov}, 303

\bibitem[{{Howe} {et~al.}(2018{\natexlab{b}}){Howe}, {Hill}, {Komm}, {Chaplin},
  {Elsworth}, {Davies}, {Schou}, \& {Thompson}}]{rachel2018}
{Howe}, R., {Hill}, F., {Komm}, R., {et~al.} 2018{\natexlab{b}}, \apjl, 862,
  L5.
\newblock \doarXiv{1807.02398}

\bibitem[{Howe {et~al.}(1999)Howe, Komm, \& Hill}]{rachel1999}
Howe, R., Komm, R., \& Hill, F. 1999, \apj, 524, 1084

\bibitem[{{Howe} {et~al.}(2002){Howe}, {Komm}, \& {Hill}}]{howe2002}
{Howe}, R., {Komm}, R.~W., \& {Hill}, F. 2002, \apj, 580, 1172,
  \dodoi{10.1086/343892}

\bibitem[{{Jain} {et~al.}(2009){Jain}, {Tripathy}, \& {Hill}}]{kiran}
{Jain}, K., {Tripathy}, S.~C., \& {Hill}, F. 2009, \apj, 695, 1567,
  \dodoi{10.1088/0004-637X/695/2/1567}

\bibitem[{{Kosovichev}(1996)}]{kosovichev}
{Kosovichev}, A.~G. 1996, \apjl, 469, L61

\bibitem[{{Kosovichev} \& {Pipin}(2019)}]{sasha2019}
{Kosovichev}, A.~G., \& {Pipin}, V.~V. 2019, \apj, 871, L20

\bibitem[{{Libbrecht} \& {Woodard}(1990)}]{libbrecht1990}
{Libbrecht}, K.~G., \& {Woodard}, M.~F. 1990, \nat, 345, 779,
  \dodoi{10.1038/345779a0}

\bibitem[{{Nishizawa} \& {Shibahashi}(1995)}]{nishizawa}
{Nishizawa}, Y., \& {Shibahashi}, H. 1995, in Astronomical Society of the
  Pacific Conference Series, Vol.~76, GONG 1994. Helio- and Astro-Seismology
  from the Earth and Space, ed. R.~K. {Ulrich}, J.~{Rhodes}, E.~J., \&
  W.~{Dappen}, 280

\bibitem[{{Palle} {et~al.}(1989){Palle}, {Regulo}, \& {Roca Cortes}}]{pere}
{Palle}, P.~L., {Regulo}, C., \& {Roca Cortes}, T. 1989, \aap, 224, 253

\bibitem[{{Pijpers} \& {Thompson}(1992)}]{sola}
{Pijpers}, F.~P., \& {Thompson}, M.~J. 1992, \aap, 262, L33

\bibitem[{{Pijpers} \& {Thompson}(1994)}]{derivative}
---. 1994, \aap, 281, 231

\bibitem[{{Scherrer} {et~al.}(1995){Scherrer}, {Bogart}, {Bush}, {Hoeksema},
  {Kosovichev}, {Schou}, {Rosenberg}, {Springer}, {Tarbell}, {Title},
  {Wolfson}, {Zayer}, \& {MDI Engineering Team}}]{mdi}
{Scherrer}, P.~H., {Bogart}, R.~S., {Bush}, R.~I., {et~al.} 1995, \solphys,
  162, 129

\bibitem[{{Scherrer} {et~al.}(2012){Scherrer}, {Schou}, {Bush}, {Kosovichev},
  {Bogart}, {Hoeksema}, {Liu}, {Duvall}, {Zhao}, {Title}, {Schrijver},
  {Tarbell}, \& {Tomczyk}}]{hmi}
{Scherrer}, P.~H., {Schou}, J., {Bush}, R.~I., {et~al.} 2012, \solphys, 275,
  207

\bibitem[{Tapping(2013)}]{tapping2}
Tapping, K.~F. 2013, Space Weather, 11, 394

\bibitem[{{Tapping} \& {Morton}(2013)}]{tapping}
{Tapping}, K.~F., \& {Morton}, D.~C. 2013, in Journal of Physics Conference
  Series, Vol. 440, Journal of Physics Conference Series, 012039

\bibitem[{{Watson} \& {Basu}(2020)}]{courtney}
{Watson}, C.~B., \& {Basu}, S. 2020, \apjl, 903, L29,
  \dodoi{10.3847/2041-8213/abc348}

\bibitem[{{Woodard} {et~al.}(1991){Woodard}, {Kuhn}, {Murray}, \&
  {Libbrecht}}]{woodard1991ApJ}
{Woodard}, M.~F., {Kuhn}, J.~R., {Murray}, N., \& {Libbrecht}, K.~G. 1991,
  \apjl, 373, L81, \dodoi{10.1086/186056}

\bibitem[{{Woodard} \& {Noyes}(1985)}]{woodard}
{Woodard}, M.~F., \& {Noyes}, R.~W. 1985, \nat, 318, 449

\end{thebibliography}

\end{document}